\documentclass[multphys,vecphys]{svmult}

\usepackage{makeidx}         
\usepackage{graphicx}        
\usepackage{multicol}        
\usepackage[bottom]{footmisc}

\makeindex             

\begin{document}

\title*{Resolving Extragalactic Star Clusters with HST/ACS}


\author{S{\o}ren S. Larsen
}


\institute{Astronomical Institute, Utrecht University, Princetonplein 5,
NL-3584 CC, Utrecht, The Netherlands, 
\texttt{larsen@astro.uu.nl}
}
%
%
\maketitle

\begin{abstract}

With HST, colour-magnitude diagrams (CMDs) can be obtained for young star 
clusters well beyond the Local Group.  Such data can help constrain 
cluster ages and metallicities, and also provide a reference against 
which intermediate- and high mass stellar models can be compared.  Here, 
CMDs are presented for two massive ($>10^5 M_\odot$) clusters and compared 
with Padua and Geneva isochrones.  The problem of the ratio of blue to red 
supergiants is also addressed. 
\end{abstract}


Star clusters remain the best approximation provided by Nature to 
``simple stellar populations'' and have a long history 
as important test labs for models of stellar evolution.  The rich old
globular clusters (GCs) in the Milky Way have played a vital role for
testing and calibrating models for low-mass stars \cite{rfp88}. However,
observational tests of intermediate- and high mass stars 
remain more scarce. The problem is largely one of statistics:
for any realistic stellar initial mass function (IMF), massive stars
constitute only a small fraction of the total number of stars in a
cluster, and have the shortest life times.  Where evolved stars are
concerned, this difficulty becomes especially acute:
a $10^5$ $M_\odot$ cluster with an age of $10^7$ years and
a Kroupa IMF \cite{kroupa02} is only expected to contain 
about 25 red supergiants, while a 10 Gyr old GC of the same mass
contains $\sim1000$ post-main sequence stars.
Clearly, evolved massive stars are rare in typical open 
clusters, and even within the entire Local Group the number of
potential targets is limited if one wishes to obtain useful constraints
on quantities such as the ratio of blue to red supergiants.

Fortunately, several galaxies within distances of a 
few Mpc contain significant populations of young ``massive'' clusters
(YMCs) with masses in excess of $10^5$ M$_\odot$ \cite{lr99}. In such
clusters, the late stages of stellar evolution start to be reasonably well 
sampled even for young ($\sim10^7-10^8$ years) ages. 
With HST, colour-magnitude diagrams (CMDs) are within reach for
some of these clusters, although care has to be taken in order to
deal with the severe crowding.  In the following, 
early results for two illustrative cases are presented.

\section{Case 1: A 35--50 Myr old cluster in NGC~1313}

The Magellanic-type galaxy NGC~1313 ($D\sim4.1$ Mpc) is known from 
ground-based imaging to host several YMCs \cite{lr99}. The western part
of the galaxy appears to have experienced a recent burst of star formation
during which also a particularly massive star cluster was formed (\#379 in 
the list of \cite{lr99}). The cluster has an age of $\sim50$ Myrs
(see below), consistent with the peak of field star formation activity in the
region \cite{lar07}. The 
integrated magnitude $M_V = -10.9$ corresponds to a mass of 
175000 $M_\odot$ for a Chabrier IMF \cite{chab03,bc03}.

\begin{figure}
\centering
\includegraphics[width=12cm]{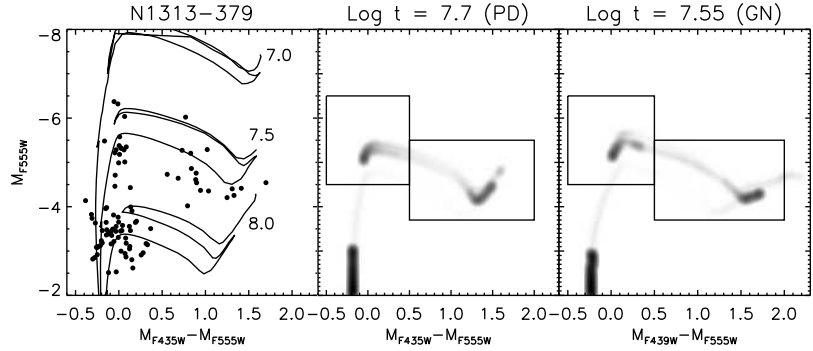}
%
%
\caption{Left: observed colour-magnitude diagram for a massive star
cluster in NGC~1313. Centre and right: Hess diagrams showing the
density of stars predicted by isochrones from the Padua 
(centre) and Geneva groups (right).  Boxes indicate regions where
red and blue supergiants are counted, as explained in the text.}
\label{fig:n1313}       
\end{figure}

The cluster is well resolved on our ACS/WFC images (Progr.\ ID 9774; 
P.I.\ S.\ S. Larsen), and the CMD
is shown in Fig.~\ref{fig:n1313} (left).  Several 
distinct features are seen: The main sequence turn-off (MSTO) is at
$m_{\rm F435} - m_{\rm F555W} \approx 0$ and $M_V \approx -3.5$.
At brighter
absolute magnitudes ($M_V\sim-5.5$) are the blue core He burning stars
(``blue supergiants'', BSGs).
The red core He burning stars (``red supergiants'', RSGs) are seen
at $m_{\rm F435} - m_{\rm F555W} > 0.5$ and $M_V\sim -4.5$. Also
shown are Padua isochrones \cite{gir02} for $Z=0.008$ and 
$\log t = 7.0$, 7.5 and 8.0. The cluster appears slightly older than 
$\log t=7.5$. 

Isochrones do not fully illustrate the distribution of stars
in the CMD. In particular, the gap between the MSTO
and the BSGs is only revealed when the \emph{density}
of stars in the CMD is shown.  The centre panel shows the
synthetic Hess diagram for a $\log t = 7.7$ Padua isochrone, while
the right-hand panel shows a $\log t = 7.55$ Geneva isochrone.
These ages correspond to MSTO masses of 7.4 $M_\odot$ and
8.1 $M_\odot$, respectively.  Both sets of models reproduce
the overall distribution of stars in the CMD fairly well, albeit for somewhat
different ages. The RSGs may be slightly too red (cool) in both cases.
The apparent difference in the model colours of the RSGs is
mostly due to the fact that WFPC2 F439W$-$F555W colours are shown for
the Geneva isochrones, while the proper ACS/WFC F435W$-$F555W colours are 
used for the Padua isochrones. 
In reality, the difference in $T_{\rm eff}$ between the Padua and Geneva 
RSGs is small, with the Geneva RSGs being some 150 K cooler.

One long-standing problem concerns the relative fraction of RSGs and BSGs
\cite{lm95,egg02}. Models tend to predict a decreasing BSG/RSG ratio
with increasing metallicity, while observations show the opposite.
Fortunately, a comparison with the data is possible without an
exact calibration between fundamental stellar properties ($L$, $T_{\rm eff}$)
and colours. 
The boxes drawn in Fig.~\ref{fig:n1313} indicate regions of the CMD where
we count RSGs and BSGs. In the observed CMD we find 16 BSGs and 17 RSGs,
corresponding to a ratio of BSG/RSG = $0.94\pm0.33$. Within the error, this is 
consistent with the model predictions:
BSG/RSG = 0.73 and 0.74 for the Padua and Geneva models.

To summarize, the agreement between models and observations seems
rather satisfactory for this cluster. This is true both for the
colours and luminosities of various types of stars and the BSG/RSG ratio.
Since the stars in this cluster only barely qualify for the 
label ``massive'', this was perhaps to be expected. It should be noted,
however, that the CMD-based age could be either 50 Myr or
35 Myr, depending on the choice of models. An important corollary
is that absolute ages derived from \emph{integrated} colours (or
spectra) are likely to be uncertain by at least the same amount.

\section{Case 2: Cluster NGC~1569-B}

The dwarf irregular galaxy NGC~1569 is well known for hosting two very 
compact, massive young clusters. Although it is about a 
factor of two closer than NGC~1313, observations of the clusters in NGC~1569
are challenging due to both their more compact structure, the general
degree of crowding, and the significant Galactic foreground
reddening ($b = 11^\circ$). Nevertheless, both clusters A and B are
resolved into individual stars in ACS/HRC data (Progr.\ ID.\ 9300; 
P.I.\ H.\ Ford); here we concentrate
on cluster B for which we also have Keck/NIRSPEC near-infrared spectroscopy 
\cite{lar07b}. The reddening-corrected absolute $V$ magnitude is $M_V=-12.2$
\cite{origlia2001}, corresponding to a mass of about 280000 $M_\odot$
for an age of 15 Myrs.

\begin{figure}
\centering
\includegraphics[width=12cm]{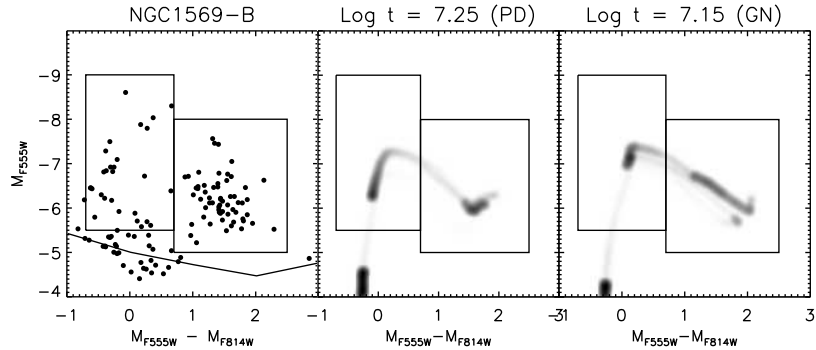}
\caption{As Fig.~\ref{fig:n1313}, but for NGC~1569-B. The 75\% completeness
limit is indicated in the left-hand panel.}
\label{fig:n1569}       
\end{figure}

The CMD for NGC~1569-B is shown in Fig.~\ref{fig:n1569}.  The
RSGs can again be clearly discerned, but the identification of
BSGs is less obvious and the MSTO is too faint to
be detected. The best fitting Padua models have an age of
$\log t = 7.25$ while the best fitting Geneva models
are slightly younger; $\log t = 7.15$. The corresponding MSTO
masses are 13 $M_\odot$ and 14 $M_\odot$,
respectively.  We have again assumed $Z=0.008$ models, 
which give the best fits, although this may not exactly match the
actual metallicity of NGC~1569.
However, neither set of models provides a
very good match to the observations for any
metallicity. Even for fairly generous limits in the selection
of BSGs, the observed BSG/RSG ratio of 28/60 = $0.47\pm0.11$ is
well below the model predictions (BSG/RSG = 1.28 and
0.87 for the Padua and Geneva models). The CMD also shows a
small number of very bright ($M_V<-8$), blue stars which are not 
predicted by any model for this age. These are redder
than the fainter blue stars, and thus unlikely to be
simple blends. One possibility is that
these stars have somewhat younger ages than the majority of stars
in cluster B.

\section{Concluding remarks}

The two cases discussed here demonstrate that useful information can be 
extracted from the CMDs of star clusters well beyond the Local Group.  
Young, massive ($>10^5$ $M_\odot$) clusters offer significant samples of 
stars with similar age and chemical composition, offering a potentially 
powerful way to test models for massive stars. Most current alternatives
either suffer from small number statistics (e.g.\ open clusters in
the Milky Way) or from difficulties disentangling the contributions
from stars of different ages, hence masses (e.g.\ resolved stellar populations
in Local Group galaxies). 

%
%
%



\printindex
\end{document}